\documentclass[twocolumn]{article} 
\newcommand{\RRR}{$I\hspace{-0.25em}R^3$}
\newcommand{\ket}[1]{\vert#1\rangle} 
\newcommand{\bra}[1]{\langle#1\vert} 
\newcommand{\bea}{\begin{eqnarray}} 
\newcommand{\eea}{\end{eqnarray}} 
\newcommand{\be}{\begin{equation}} 
\newcommand{\ee}{\end{equation}} 
\newcommand{\bdm}{\begin{displaymath}} 
\newcommand{\edm}{\end{displaymath}} 
\newcommand{\bq}{\begin{quote}} 
\newcommand{\eq}{\end{quote}} 
\newcommand{\medno}{\par\medskip\noindent} 
\newcommand{\braket}[2]{\langle#1\vert#2\rangle}

\begin{document}
\title{THE WORLD ACCORDING TO QUANTUM MECHANICS\\
(OR, THE 18 ERRORS OF HENRY P. STAPP)}
\author{Ulrich Mohrhoff\\ 
Sri Aurobindo International Centre of Education\\ 
Pondicherry-605002 India\\ 
\normalsize\tt ujm@satyam.net.in} 
\date{}
\maketitle 
\begin{abstract}
Several errors in Stapp's interpretation of quantum mechanics and its application to 
mental causation (Henry P. Stapp, ``Quantum theory and the role of mind in nature,'' 
e-Print quant-ph/0103043, to appear in Foundations of Physics) are pointed out. An 
interpretation of (standard) QM that avoids these errors is presented.
\end{abstract}

\section{\large INTRODUCTION}

According to some 
theorists~\cite{LonBau,vN,Wigner,Heisbg,Page,Squires,Lockwood,Albert}, 
consciousness is needed for making sense of quantum mechanics (QM). According to 
others~\cite{JY,Penrose,GGG,Hameroff,Eccles}, QM is needed in order to understand 
consciousness and/or account for its causal efficacy. In a recent contribution to this 
journal, Henry P. Stapp~\cite{Stapp} argues for both: Consciousness is essential for 
understanding QM, and QM is essential for the causal efficacy of consciousness. In the 
present article I point out a number of errors that mar Stapp's theory---inconsistencies, 
fallacies, and conclusions founded on such logical mistakes,---and offer an 
interpretation of (standard) QM that avoids these errors. In addition I outline an 
alternative account of mental causation~\cite{UMIEC,UMPI}.

Section 2 stresses the fact that an algorithm for assigning probabilities to the possible 
results of possible measurements cannot also represent an evolving state of affairs 
(Stapp's first error, ${\cal E}_1$). The introduction of consciousness into discussions of 
QM (${\cal E}_2$) serves no other purpose than to provide gratuitous solutions to a 
pseudo-problem arising from ${\cal E}_1$. Stapp's third error (${\cal E}_3$), pointed 
out in Sec.~3, is a category mistake. It consists in his treating possibilities as if they 
possessed an actuality of their own. This leads to the erroneous notion that possibilities 
are things (``propensities'') that exist and evolve in time (${\cal E}_4$).

Stapp offers two arguments purporting to support the existence of a dynamically 
preferred family of constant-time hypersurfaces. The first invokes astronomical data, 
which support the existence of a historically preferred family of hypersurfaces but not of 
a dynamically preferred one (${\cal E}_5$). The second is fallacious because it involves 
inconsistent combinations of (individually valid) counterfactuals (${\cal E}_6$). Equally 
fallacious, therefore, is Stapp's ``proof'' of the occurrence of faster-than-light transfers 
of information (${\cal E}_7$). This is discussed in Sec.~4.

According to Stapp, granting free will to experimenters leads to a physical reality 
inconsistent with the ``block universe'' of special relativity (SR), a reality that unfolds in 
response to choices. This is one (${\cal E}_8$) of a cluster of misconceptions arising from 
the erroneous notion that the experiential now, and the temporal distinctions that we 
base on it, have anything to do with the physical world (${\cal E}_9$). Objectively, the 
past, the present, and the future exist in exactly the same atemporal sense. There is no 
such thing as ``an evolving objective physical world'' (${\cal E}_{10}$), and there is no 
such thing as an objectively open future or an objectively closed past (${\cal E}_{11}$). 
The results of performed measurements are always ``fixed and settled.'' What is 
objectively open is the results of unperformed measurements. This is discussed in Sec.~5.

The subject of causation is broached in Sec.~6. Causality, as Hume~\cite{Hume} 
discovered two and a half centuries ago, lies in the eye of the beholder. While classical 
physics permits the anthropomorphic projection of causality into the physical world 
with some measure of consistency, quantum physics does not. Trying to causally explain 
the quantum-mechanical correlations is putting the cart in front of the horse. It is the 
correlations that explain why causal explanations work to the extent they do. Stapp's 
attempt to involve causality at a more fundamental level (${\cal E}_{12}$) depends 
crucially on his erroneous view that the factual basis on which quantum-mechanical 
probabilities are to be assigned is determined by Nature rather than by us (${\cal 
E}_{13}$).

Sections 7 to 9 are of a more constructive nature. In Sec.~7 the significance of the 
existence of objective probabilities is discussed. They betoken an objective indefiniteness 
that is crucial for the stability of matter, and they imply the extrinsic nature of the 
values of quantum-mechanical observables. Extrinsic properties presuppose intrinsic 
ones, and this implies the co-existence of two logically distinct domains. How these are 
related to each other is discussed in Sec.~8. Section~9 gets to the heart of QM, which 
concerns the spatiotemporal differentiation of reality. The fact that this is finite makes 
QM as inconsistent with a fundamental assumption of field theory as SR is with absolute 
simultaneity. Stapp shares this erroneous assumption when he considers the physical 
world differentiated into ``neighboring localized microscopic elements'' (${\cal E}_{14}$).

Section 10 begins by observing that, contrary to Stapp's contention (${\cal E}_{15}$), 
the freedom to choose is a classical phenomenon. Subsequently the reader is taken 
through the steps of Stapp's account of mental causation, and a number of further 
errors (not all enumerated) are pointed out, such as: The objective brain can 
(sometimes) be described as a decoherent mixture of ``classically described brains'' all of 
which must be regarded as real (${\cal E}_{16}$). Crucial to Stapp's account is the 
metaphor of the experimenter as interrogator of Nature. Within the Copenhagen 
framework, which accords a special status to measuring instruments, this is a fitting 
metaphor for a well-defined scenario. In Stapp's framework, which accords a special 
status to the neural correlates of mental states, it is not (${\cal E}_{17}$). Its sole purpose 
is to gloss over the disparity between physical experimentation and psychological 
attention. Once this purpose is achieved, the metaphor is discarded, for in the end 
Nature not only provides the answers but also asks the questions. The theory Stapp ends 
up formulating is completely different from the theory he initially professes to formulate 
(${\cal E}_{18}$), for in the beginning consciousness is responsible for state vector 
reductions, while in the end a new physical law is responsible, a law that in no wise 
depends on the presence of consciousness.

The final section contains concluding remarks.

\section{\large QUANTUM MECHANICS AND\\
CONSCIOUSNESS}
\label{qmc}
Stapp capitalizes on von Neumann's formulation and interpretation of QM as a theory 
of the objective world interacting with human consciousness~\cite{vN}. The unobserved 
world evolves according to a dynamical equation such as the Schr\"odinger equation, 
while observations cause ``a sudden change that brings the objective physical state of a 
system in line with a subjectively felt psychical reality''~\cite{Stapp}. This makes QM 
``intrinsically a theory of mind--matter interaction,'' and more specifically a theory 
``about the mind--brain connection''~\cite{Stapp}. Each of those sudden changes
\bq
injects one ``bit'' of information into the quantum universe. These bits are stored in the 
evolving objective quantum state of the universe, which is a compendium of these bits of 
information\ldots. Thus the quantum state has an ontological character that is in part 
matter like, since it is expressed in terms of the variables of atomic physics, and it 
evolves between events under the control of the laws of atomic physics. However, each 
event injects the information associated with a subjective perception by some observing 
system into the objective state of the universe~\cite{Stapp}.
\eq
Whatever else a quantum state may represent, there can be no doubt that it is first of all 
an algorithm for assigning probabilities to the possible results of possible measurements. 
This is evident from the minimal instrumentalist interpretation of QM, the common 
denominator of all possible interpretations~\cite{Redhead}. It is also evident from J.M. 
Jauch's definition of the ``state'' of a quantum-mechanical system as a probability 
measure resulting from a preparation of the system and his proof~\cite{Jauch}---based 
on Gleason's theorem~\cite{Gleason}---that every such probability measure has the 
well-known density-operator form, which reduces to the familiar Born probability 
measure if the density operator is idempotent. But if a quantum state is a probability 
algorithm, then it cannot also represent an actual state of affairs. How could it? A state 
of affairs is an entirely different kettle of fish; it falls under an entirely different 
category. This immediately disposes of the ``measurement problem'' in its crudest form, 
which arises only if state vectors or density operators are regarded as representing 
actual states of affairs.

The quantum revolution was guided by the vision of Niels Bohr. In 1913, Bohr rejected 
classical physics wholesale, initiated the creation of an entirely new physics, and rallied 
physicists to complete it. The same genius later impressed upon physicists the true 
import of the new physics: {\it QM spells the end of mathematical realism.} The 
symbols and formulae of the new physics can no longer be interpreted as mirroring 
(representing, describing) the physical world. All that QM places at our disposal is 
probability measures; it assigns probabilities to the possible outcomes of 
possible measurements. 
Any attempt to go beyond the statistical regularities encapsulated by the laws of QM 
must, at the very least, be consistent with the incontestable probabilistic significance of 
quantum states. Interpretations that grant quantum states the ontological significance of 
a state of affairs {\it do not satisfy this fundamental requirement}.

According to John Bell~\cite{Bell}, ``measurement'' is a bad word. The really bad word, 
however, is ``state,'' owing to the obvious suggestion that a quantum state represents a 
state in the usual sense of the word. ``Measurement'' isn't a good word either, but 
this is more easily repaired. QM represents the contingent properties of a physical 
system $S$---the properties that $S$ can but does not necessarily possess---as subspaces 
of some vector space. Bohr rightly insisted that a contingent property $q$ can be 
attributed to $S$ only if the following criterion is satisfied: The possession of $q$ by $S$ 
must be {\it indicated}. The properties of quantum systems are {\it extrinsic} in the 
specific sense that they cannot be attributed unless they are 
indicated~\cite{Mohrhoff00}. {\it No property is a possessed property unless it is an 
indicated property.} Since the properties of quantum systems are usually indicated by 
what we call ``measuring devices,'' this creates the absurd impression that laboratory 
instruments play a fundamental ontological role. What actually plays a fundamental role 
in the formulation of QM is {\it facts}---actual states of affairs---and especially such facts 
as indicate the possessed properties of quantum-mechanical systems. The word that 
ought to replace ``measurement'' in any ontological interpretation of QM is 
``(property-indicating) fact.''

What is a fact? {\it The Concise Oxford Dictionary} (8th edition, 1990) defines ``fact'' as 
a thing that is {\it known} to have occurred, to exist, or to be true; a datum of {\it 
experience}; an item of verified {\it information}; a piece of {\it evidence}. Other 
dictionaries give variations on the same theme. Should we conclude from this that the 
editors of dictionaries are idealists wanting to convince us that the existence of facts 
presupposes knowledge or experience? Obviously not. The correct conclusion is that 
``fact,'' like ``existence,'' like ``reality,'' is so fundamental a concept that it simply {\it 
cannot be defined}. So what is the editor of a dictionary to do? The obvious thing is to 
fall back on the metalanguage of epistemology. Which is precisely what Bohr did to 
bring home to lesser spirits the extrinsic nature of the properties of quantum systems.

If ``fact'' is so fundamental a term that it cannot be defined, the existence of facts---the 
factuality of events or states of affairs---cannot be accounted for, any more than we can 
explain why there is anything at all, rather than nothing. (If something can be accounted 
for, it can be defined in terms of whatever accounts for it.) Before the mystery of 
existence---the existence of {\it facts}---we are left with nothing but sheer 
dumbfoundment. Any attempt to explain the emergence of facts (``the emergence of 
classicality,'' as it is sometimes called) must therefore be a wholly gratuitous endeavor.

Classical physics deals with nomologically possible worlds---worlds consistent with 
physical theory. It does not uniquely determine the actual world. Identifying the actual 
world among all nomologically possible worlds is strictly a matter of 
observation. Does this imply that classical physics presupposes conscious observers? 
Obviously not. In classical physics the actual course of events is in principle fully 
determined by the actual initial conditions (or the actual initial and final conditions). In 
quantum physics it also depends on unpredictable actual states of affairs at later (or 
intermediate) times. Accordingly, picking out the actual world from all nomologically 
possible worlds requires observation not only of the actual initial conditions (or the 
actual initial and final conditions) but also of those unpredictable actual states of affairs. 
Does this imply that quantum physics presupposes conscious observers? If the answer is 
negative for classical physics, it is equally negative for quantum physics.

QM concerns statistical correlations between (observer-independent) facts, and these 
correlations warrant interpreting the facts as indicative of properties. That is, they 
warrant the existence of a physical system to which the indicated properties can be 
attributed. Suppose that we perform a series of position measurements, and that every 
position measurement yields exactly one result (that is, each time exactly one detector 
clicks). Then we are entitled to infer the existence of a persistent entity, to think of the 
clicks given off by the detectors as matters of fact about the successive positions of this 
entity, to think of the behavior of the detectors as position measurements, and to think 
of the detectors as detectors. If instead each time exactly two detectors click, we are 
entitled to infer the existence of two entities or, rather, of a physical system with the 
property of having two components. This property is as extrinsic as are the measured 
positions. There is a determinate number of entities only {\it because} every time the 
same number of detectors click. Not only the properties of things but also the number of 
existing things supervenes on the facts.

The ontological dependence of the properties and the number of existing things on facts 
warrants a distinction between two domains, a {\it classical domain} of facts and a {\it 
quantum domain} of properties that supervene on the facts. Owing to the ontological 
dependence of the domain of indicated properties on the domain of property-indicating 
facts, the quantum domain cannot account for the existence of the classical domain. If 
one nevertheless assumes that (i)~the ultimate physical reality is the quantum domain, 
and that (ii)~the existence of the domain of facts can, and therefore should, be accounted 
for, then consciousness becomes an obvious candidate: Facts exist because they are 
perceived.

If in addition one wrongly assumes with von Neumann that the quantum state 
represents the dynamical evolution of the quantum domain, one is faced with the 
spurious problem of reconciling two disparate modes of evolution. The obvious 
``solution'' then is to consider one mode of evolution intrinsic to the quantum domain 
and to blame the other mode on the intervention of conscious observers. This, in brief, is 
how some of the brightest physicists were led to conclude that QM is an epistemic 
theory, concerned with our knowledge or experience of the factual situation rather than 
the factual situation itself. One is reminded of the various God-of-the-gaps proposals of 
the past. While invoking divine intervention as a filler for explanatory gaps is no longer 
in fashion, there is a tendency to invoke consciousness instead.

For reasons that are obvious rather than mysterious, probability measures have two 
modes of ``change.'' Probabilities are assigned on the basis of relevant facts, and they 
are assigned to sets of possible property-indicating facts, or else to the properties 
indicated by such facts. They depend (i)~on the time $t$ of the possible facts to which 
they are assigned and (ii)~on the facts on which the assignment is based. They therefore 
``change'' not only in a ``deterministic'' manner, as functions of $t$, but also 
unpredictably with every fresh relevant fact. The successful completion of a 
measurement is the relevant fact {\it par excellence}. If the outcome is unpredictable, as 
it generally is, it has to be included among the relevant facts on which further 
probability assignments ought to be based. The outcome being unpredictable, the basis 
of relevant facts changes unpredictably as a matter of course, and so do the probabilities 
assigned on this basis.

The reason for the quotation marks is that a probability is not the kind of thing that {\it 
changes}. To see this, consider the Born probability $p\,(R,t)$ of finding a particle in a 
region $R$ at a time~$t$. While few would think of this probability as something that 
exists inside~$R$, many appear to think of it as something that exists at the time~$t$. 
The prevalent idea is that the possibility of finding the particle inside $R$ exists at all 
times for which $p\,(R,t)>0$, so the probability associated with this possibility also exists 
at all those times and changes as a function of time. Yet the possibility that a 
property-indicating state of affairs obtains at the time $t$ is not something that exists at 
the time~$t$, anymore than the possibility of finding the particle in $R$ is something 
that one can find inside~$R$. And the same obviously holds true of the probability 
associated with this possibility. $p\,(R,t)$ isn't something of which we can say {\it when} 
it exists. {\it A fortiori} it isn't something that can change or evolve. All 
quantum-mechanical probability assignments are conditional on the existence of a 
matter of fact about the value of a {\it given} observable at a {\it given} time. $p\,(R,t)$ 
is not associated with the possibility that all of a sudden, at the time~$t$, the particle 
``materializes'' inside~$R$. It is the probability with which the particle is found in $R$, 
{\it given} that at the time $t$ it is found in one of a set of mutually disjoint regions (no 
matter which one, $R$ being one of them). The parameter $t$ on which this probability 
depends is the specified time of this actually or counterfactually performed position 
measurement. It refers to the time of a position-indicating state of affairs, the existence 
of which is {\it assumed}.

So much for the ``change'' of $p\,(R,t)$ associated with the argument~$t$. It is obvious 
that the sense in which $p\,(R,t)$---for a {\it fixed} value of $t$---``changes'' when 
assigned on the basis of a fresh set of facts, is also not the sense in which a state of affairs 
evolves as time passes.

By misconstruing quantum states as evolving states of affairs with two modes of change, 
von Neumann~\cite{vN} created a number of pseudo-problems and spawned an entire 
industry dedicated to generating gratuitous solutions. Are wave function collapses in the 
mind but not in the world? Or are they in the mind {\it because} they are in the world? 
Or are they in the mind and {\it therefore} in the world? The first option leads to the 
many-worlds (or many-minds~\cite{Albert}) extravaganza, the second to nonlinear 
adulterations of QM~\cite{GRW,Pearle,BG}, the third to epistemic interpretations. If 
the premise is that system $S$ enters into a ``state of entanglement'' with apparatus $A$ 
then apparatus $A$ enters into a ``state of entanglement'' with Cecily's brain as she 
takes cognizance of the measurement outcome. The definiteness of observation reports 
combined with the principle of psycho-physical parallelism~\cite{vN}---subjective 
perceptions correspond to objective goings-on in the brain---then spells collapse. Hence 
Cecily's perceptions exert a causal influence on the ``objective physical 
state''~\cite{Stapp} of $S$ via the objective goings-on in her brain and their 
entanglement with~$S$.

The bottom line: The introduction of consciousness into interpretations of QM affords 
nothing but gratuitous solutions to pseudo-problems. These pseudo-problems arise 
whenever quantum states are construed, inconsistently with the statistical significance of 
quantum states, as evolving states of affairs. There does exist an extra-theoretical 
element that cannot be accounted for by either classical or quantum physics, but this 
neither is consciousness nor can be accounted for in epistemic terms. It is the actuality of 
exactly one of all nomologically possible worlds, or the factuality of property-indicating 
facts. This (f)actuality owes nothing to the consciousness of an ``observing 
system''~\cite{Stapp}; it is what distinguishes the world from any theory about the 
world.

\section{\large QUANTUM MECHANICS AND\\
PROPENSITIES}

The following characterizations of the physical world and of physical states are central 
to Stapp's ``objective interpretation of von Neumann's formulation of quantum 
theory''~\cite{Stapp}:
\bq
The observed physical world is described\ldots by a mathematical structure that can best 
be characterized as representing {\it information} and {\it propensities\/}: the {\it 
information} is about certain {\it events} that have occurred in the past, and the {\it 
propensities} are objective tendencies pertaining to future events.

The objective physical state is\ldots converted from a material substrate to an 
informational and dispositional substrate that carries both the information incorporated 
into it by the psychical realities, and certain dispositions for the occurrence of future 
psychical realities~\cite{Stapp}.
\eq
The possibility that something happens at the time $t$, recall, is not something that 
exists at the time~$t$, anymore than the possibility of finding a particle in a region $R$ 
is something that one can find inside~$R$. A possibility is not the kind of thing that 
persists and changes in time. To think of possibilities as if they possessed an actuality of 
their own, different from the actuality of facts, and as if they persisted and changed 
(``evolved'') in time, is an obvious category error. This logical mistake gives rise to the 
somewhat gentler avatar of the ``measurement problem,'' which asks: How is it that 
during a measurement one of the persisting possibilities (or worse, one of the changing 
probabilities associated with them%
\footnote{``Above all, we would like to understand how it is that probabilities become 
facts.''---S. Treiman~\cite{Treiman}. Dozens of similar phrases can be found in the 
literature.})
becomes a fact, while the others cease to exist? Saying in common language that a 
possibility becomes a fact is the same as saying that something that is 
possible---something that {\it can} be a fact---actually {\it is} a fact. How can that be a problem? 
This non-problem becomes a pseudo-problem if one forgets that there is only one kind of 
actuality and misconstrues the common-language ``existence'' of a possibility as a second 
kind of actuality, called ``propensity''\cite{Popper} or 
``potentiality''\cite{Heisenberg,Shimony}, that gets converted into the genuine article 
when a measurement is made.

If one wants to associate a measurement with the ``actualization of a possibility'' in a 
logically coherent manner, one must not portray it as a transition from an earlier state of 
affairs, in which the possibility ``exists'' as a possibility, to a later state of affairs, in 
which it is a fact. The possibilities to which QM assigns probabilities can all be 
formulated in the following manner:
\medno
(S)~A measurement of the (system-specific) observable $Q$ performed at the time $t$ 
yields the result $q_k$.
\medno
(Conjunctions can be formulated as well: ``Measurements of $Q_1$ and $Q_2$ 
performed at the respective times $t_1$ and $t_2$ yield the respective results $q^1_i$ 
and $q^2_k$.'') Owing to its explicit reference to the time of measurement, such a 
sentence cannot {\it become} true or false. If (S) is true, it always has been and always 
will be true, and if it is false, it always has been and always will be false. Saying that the 
possibility expressed by (S) has been actualized is therefore the same as saying that (S) is 
true. Nothing more must be read into this sentence. The question as to when this 
actualization took place is utterly meaningless. If the ``actualization of a possibility'' 
involves a ``transition,'' it is the logical transition from a possible world in which $Q$ is 
not measured at the time $t$ to the actual world in which $Q$ is successfully measured 
at the time~$t$.

The bottom line: The only problem that is addressed by the introduction of propensities 
or potentialities into interpretations of QM is another pseudo-problem originating from 
another logical mistake. It arises if one thinks of the possibilities to which QM refers, 
and of the probabilities it assigns to them, as if they constituted a self-existent matrix 
from which facts arise.

\section{\large QUANTUM MECHANICS AND\\
SPECIAL RELATIVITY}
\label{qmsr}
It is well known that the statistical regularities with which QM is concerned are 
consistent with SR, while von Neumann's interpretation of states as evolving, collapsible 
states of affairs is not. Stapp tries to reconcile SR with von Neumann's interpretation by 
giving ``special objective physical status''~\cite{Stapp} to a particular family of 
constant-time hypersurfaces: State reductions occur globally and instantaneously with 
respect to this family of hypersurfaces. He offers two arguments purporting to support 
the existence of a ``dynamically preferred sequence of instantaneous 
`nows'\,''~\cite{Stapp}. The first invokes astronomical data that, incontestably, indicate 
the existence of a (cosmologically) preferred sequence of ``nows.'' This argument fails to 
distinguish sufficiently between (i)~a {\it dynamically} preferred sequence of 
hypersurfaces---a sequence implied by the dynamical laws and present in every possible 
world consistent with them---and (ii)~a preferred sequence of contingent and historical 
character that is not implied by the dynamical laws and therefore is not a feature of 
every nomologically possible world. The astronomical data suggest the existence of a 
historically preferred sequence but give no evidence of a dynamically preferred one.

The second argument is based on an experiment of the kind first discussed by Lucien 
Hardy~\cite{Hardy}. There are two regions $L$ and $R$ in spacelike separation such 
that, in a certain coordinate system $(x,y,z,t)$, $R$ is earlier than $L$. Two two-valued 
observables $L_i$ and $R_i$ ($i=1,2$) can be measured in each region. The joint 
probabilities of the various possible results are determined by
\be
\ket{\Psi}=\ket{L1+,R1-}-\ket{L2-,R2+}\braket{L2-,R2+}{L1+,R1-},
\ee
and they warrant the following assertions:
\bea
L_1-\Rightarrow R_2+,\label{a}\\
R_2+\Rightarrow L_2+,\label{b}\\
L_2+\Rightarrow R_1-,\label{c}\\
L_1-\not\Rightarrow R_1{-}.\label{d}
\eea
In longhand (\ref{a}) says that if $L_1$ is measured and the result is $L_1-$ then if 
$R_2$ is measured the result is~$R_2+$. On account of (\ref{a}), the conditional ``If 
$R_2$ is measured then $R_2+$ is obtained'' is valid on condition that $L_1$ is 
measured and $L_1-$ is obtained. On account of (\ref{b}), the conditional ``If $L_2$ is 
measured then $L_2+$ is obtained'' is valid on condition that $R_2$ is measured and 
$R_2+$ is obtained. Combining these two conditional statements, Stapp arrives at the 
following conclusion:
\bq
(A) If $L_1$ and $R_2$ are measured and the outcome of the measurement of $L_1$ is 
$L_1-$ then if $L_2$ had been measured instead of $L_1$ the outcome $L_2+$ would 
have been obtained.
\eq
While it is legitimate to ask for the result that would have been obtained if $L_2$ had 
been measured instead of $L_1$, it is illegitimate to base the answer on the assumption 
that $L_1$ was measured. There is no quantum state such that both $L_1$ and $L_2$ 
are dispersion-free. It is therefore logically impossible to counterfactually assign a 
definite value to $L_2$ on the basis of a result of a measurement of $L_1$. Probabilities 
can be assigned counterfactually (that is, to the possible results of unperformed 
measurements), but only if the facts on the basis of which they are assigned are 
consistent with the measurements to the possible results of which they are assigned. The 
combination of the above two conditional statements is illegitimate because it assigns a 
probability to a possible outcome of a measurement of $L_2$ on the basis of an outcome 
of a measurement of $L_1$. The existence of an outcome of the former measurement is 
inconsistent with the existence of an outcome of the latter measurement.

By an equally fallacious route Stapp arrives at the conclusion that the following 
statement is false:
\bq
(B) If $L_1$ and $R_1$ are measured and the outcome of the measurement of $L_1$ is 
$L_1-$ then if $L_2$ had been measured instead of $L_1$ the outcome $L_2+$ would 
have been obtained.
\eq
Here, too, two individually correct statements are combined in a logically inconsistent 
manner: (i)~The antecedent (including the assumption that $L_1$ is measured) and 
(\ref{d}) jointly imply that sometimes $R_1+$ is obtained. (ii)~The assumption that 
$R_1+$ is obtained and (\ref{c}) jointly imply that a measurement of $L_2$ does not 
yield $L_2+$. Both statements are correct, but they consider logically incompatible 
situations, so that no valid conclusions can be drawn from their conjunction.

To buttress his conclusions, Stapp makes the assumption that in the coordinate system 
$(x,y,z,t)$ no backward-in-time influences occur. He makes this assumption to 
exclude two possibilities: (i)~the possibility that the result $R_2+$ causally depends on 
the result $L_1-$ and therefore cannot be invoked to infer that $L_2+$ would have been 
obtained had $L_2$ been measured instead of $L_1$, and (ii)~the possibility that the 
occasional result $R_1+$ causally depends on the result $L_1-$ and therefore cannot be 
invoked to infer that, whenever $R_1+$ is obtained, $L_2+$ would not have been 
obtained had $L_2$ been measured instead of $L_1$. Causal considerations, however, 
are completely irrelevant to the validity or otherwise of Stapp's conclusions. What 
invalidates his conclusions is the fact that statements concerning the value of one 
observable cannot be based on assumptions concerning the value of another observable 
if the two observables are incompatible.

From the truth of (A) (symbolically: $L_1{-}\wedge R_2\pm\Rightarrow L_2+$) and the 
falsity of (B) (symbolically: $L_1{-}\wedge R_1\pm\not\Rightarrow L_2+$) Stapp 
infers that
\bq
a theoretical constraint upon what nature can choose in region $L$, under conditions 
freely chosen by the experimenter in region $L$, depends nontrivially on which 
experiment is freely chosen by the experimenter in region $R$\ldots. But this 
dependence cannot be upheld without the information about the free choice made in 
region $R$ getting to region $L$: {\it some sort of faster-than-light transfer of 
information is required.}
\eq
The occurrence of $L_1$ on the left-hand side of the above symbolic formulae 
representing the truth of (A) and the falsity of (B), respectively, is logically inconsistent 
with the occurrence of $L_2$ on the right-hand side. Therefore Stapp's second 
argument also fails to establish the ``special objective physical status''~\cite{Stapp} of a 
particular family of constant-time hypersurfaces, or the existence of a ``dynamically 
preferred sequence of instantaneous `nows'.''

Stapp considers the nonexistence of a such a dynamically preferred sequence in SR an 
empirically unwarranted ``metaphysical idea.'' What would the {\it existence} of such a 
sequence amount to?

The great theoretical breakthrough that gave us SR came with the realization that 
(i)~what propagates with an invariant speed requires no medium, and that (ii)~the 
invariant speed is the finite speed of light~$c$. ``Invariant'' means ``independent of the 
inertial frame in which it is measured.'' It is not hard to see that there cannot be more 
than one invariant speed.

If an event $e_1$ at ($x_1,t_1$) is the cause of an event $e_2$ at ($x_2,t_2$), the fact 
that $e_2$ happens at $t_2$, rather than at any other time, has two possible 
explanations. If the causal connection is {\it mediated}, and if $x_1$, $t_1$, and $x_2$ 
are fixed, $t_2$ is determined by the speed of mediation. This could be the speed of a 
material particle traveling from $x_1$ to $x_2$ or the speed of signal propagation in an 
elastic medium. On the other hand, if the causal connection is {\it unmediated}, $t_2$ is 
determined by the metric structure with respect to which the dynamical laws are 
formulated, and this metric structure defines an invariant speed. In a nonrelativistic 
world the metric structure defines an infinite invariant speed---what propagates 
instantaneously, or with an infinite speed, in one inertial frame, does so in every other 
inertial frame,---whereas in a relativistic world it defines a finite invariant speed. In a 
nonrelativistic world, accordingly, the dynamical laws allow for an unmediated causal 
connection between $e_1$ and $e_2$ provided that, invariantly, $t_2=t_1$, while in a 
relativistic world they allow for an unmediated causal connection between $e_1$ and 
$e_2$ provided that, invariantly, $(x_2-x_1)/(t_2-t_1)=c$.

We live in a world in which the dynamical laws constrain unmediated signal 
propagation to null geodesics.%
\footnote{For reasons that are psychological rather than physical, most physicists believe 
that photons are particles that mediate influences, rather than elementary unmediated 
influences. (I am not saying that they are either.)}
Signal propagation occurring along either timelike or spacelike geodesics cannot be 
invariant and therefore cannot be unmediated. By claiming the existence of a 
dynamically preferred family of constant-time hypersurfaces, Stapp therefore effectively 
postulates the existence of a quantum counterpart to the luminiferous ether. He may be 
inclined to deny this, but as long as the speed of light is invariant, there can be no other 
explanation why state reduction occurs with respect to one family of hypersurfaces 
rather than another (assuming that it does occur). The only way for Stapp to avoid the 
implication of a ``quantum ether'' is to revert to the metric of Galilean relativity, which 
implies the existence of preferred family of constant-time hypersurfaces, at the expense 
of reintroducing the luminiferous ether. So which is the unwarranted ``metaphysical 
idea''? The existence of a preferred family of hypersurfaces, which, combined with the 
invariance of the speed of light, implies the existence of one ether or another, or the 
nonexistence of such a preferred family?

\section{\large QUANTUM MECHANICS AND\\
THE EXPERIENTIAL NOW}

We are accustomed to the idea that the redness of a ripe tomato exists in our minds, 
rather than in the physical world. We find it incomparably more difficult to accept that 
the same is true of the experiential now: It has no counterpart in the physical world. 
There simply is no objective way to characterize the present. And since the past and the 
future are defined relative to the present, they too cannot be defined in objective terms. 
The temporal modes past, present, and future can be characterized only by how they 
relate to us as conscious subjects: through memory, through the present-tense 
immediacy of qualia (introspectible properties like pink or turquoise), or through 
anticipation. In the world that is accessible to physics we may qualify events or states of 
affairs as past, present, or future {\it relative to} other events or states of affairs, but we 
cannot speak of {\it the} past, {\it the} present, or {\it the} future.

The proper view of physical reality therefore is not only what Nagel~\cite{Nagel} has 
called ``the view from nowhere'' (the objective world does not contain a preferred 
position corresponding to the spatial location whence I survey it); it is also what Price 
has called ``the view from nowhen''~\cite{Price}: The objective world does not contain a 
preferred time corresponding to the particular moment (the present) at which I 
experience it. The objective world contains spatial and temporal relations as well as the 
corresponding relata, but it does not contain any kind of basis for the distinction 
between a past, a present, and a future. The idea that some things exist not yet and 
others exist no longer is as true and as false as the idea that a ripe tomato is red.

If we conceive of temporal relations, we conceive of the corresponding relata 
simultaneously---they exist at the same time {\it in our minds}---even though they 
happen or obtain at different times in the objective world. Since we can't help it, that 
has to be OK. But it is definitely not OK if we sneak into our simultaneous spatial 
mental picture of a spatiotemporal whole anything that advances across this 
spatiotemporal whole. We cannot mentally represent a spatiotemporal whole as a 
simultaneous spatial whole and then imagine this simultaneous spatial whole as 
persisting in time and the present as advancing through it. There is only one time, the 
fourth dimension of the spatiotemporal whole. There is not another time in which this 
spatiotemporal whole persists as a spatial whole and in which the present advances, or in 
which an objective instantaneous state {\it evolves}. If the present is anywhere in the 
spatiotemporal whole, it is trivially and vacuously everywhere---or, rather, everywhen.

In a world that has no room for an advancing now, time does not ``flow'' or ``pass.'' 
Objective time is a set of temporal relations between temporal relata that owe their 
successive character to our minds, rather than to anything in the objective world. To 
philosophers the perplexities and absurdities entailed by the notion of an objective 
advancing present or an objectively flowing time are well known~\cite{time}. Physicists 
began to recognize the subjectivity of the present and the nonexistence of an evolving 
instantaneous state with the discovery of the relativity of simultaneity. In the 
well-known words of Hermann Weyl~\cite{Weyl},
\bq
The objective world simply {\it is\/}; it does not {\it happen}. Only to the gaze of my 
consciousness, crawling upward along the life line of my body, does a section of this 
world come to life as a fleeting image in space which continuously changes in time.
\eq
Not only the fleeting and continuously changing image but also the ``crawling upward'' 
of our consciousness is a subjective phenomenon. So is change, and so is becoming.

The myth of an evolving instantaneous physical state is supported by a folk tale of 
considerable appeal. It goes like this: Since the past is no longer real, it can influence the 
present only through the mediation of something that persists through time. Causal 
influences reach from the nonexistent past into the nonexistent future by being ``carried 
through time'' by something that ``stays in the present.'' There is, accordingly, an 
evolving instantaneous state, and this includes not only all presently possessed properties 
but also everything in the past that is causally relevant to the future. This is how we 
come to conceive of ``fields of force'' that evolve in time (and therefore, in a relativistic 
world, according to the principle of local action), and that ``mediate'' between the past 
and the future (and therefore, in a relativistic world, between local causes and their 
distant effects). It is also how we come to believe that the state vector plays a similar 
causally mediating role. It is high time that we outgrow these incoherent beliefs.

There was a time when characterizing something as subjective was tantamount to 
denigrating it as an ``illusion.'' This is not my intention. Change and becoming are no 
less real and no less significant for being subjective. My point is that a theory of 
mind--matter interaction, which is what Stapp's theory purports to be, must proceed 
from a clear understanding of which features of the experienced, phenomenal world do, 
and which do not, have a counterpart in the physical world. Instead of proceeding from 
such an understanding, Stapp's theory features the chimera of ``an evolving objective 
physical world.'' It is ``a theory of the interaction between the evolving objective state of 
the physical universe and a sequence of mental events, each of which is associated with a 
localized individual system.''

It might be argued that without an objective becoming there can be no freedom of 
choice. Such freedom seems to require an open future. In fact, the apparent 
incompatibility of free volition with the ``block universe'' of SR is one of the reasons why 
Stapp rejects the idea that ``all of history can be conceived to be laid out in a 
four-dimensional spacetime.'' According to him, free will ``leads to a picture of a reality 
that gradually unfolds in response to choices that are not necessarily fixed by the prior 
physical part of reality alone.''

This too is an error. All that the freedom to choose requires is the impossibility of {\it 
knowing}, at any one time, what is the case at a later time. Obviously, if I can know at 
the time $t$ a state of affairs that obtains at the time $t'>t$, I cannot conceive of it as 
causally dependent on a free choice made by me in the interval between $t$ and~$t'$. If 
at any time $t$ I can know the future (relative to $t$), I cannot conceive of myself as a 
free agent. On the other hand, if the possibility of foreknowledge does not exist, I cannot 
merely conceive of myself as a free agent. I can actually be a free agent, for the only 
thing that logically prevents me from being responsible for a later state of affairs is the 
possibility of knowing the same before the relevant choices are made. The fact that the 
future in a sense ``already'' exists is no reason why choices made by me at earlier times 
cannot be partly responsible for it. (The future relative to a time $t$ exists ``already'' not 
in the sense that it exists simultaneously with what exists at the time $t$, which would be 
self-contradictory, but in the sense that it exists {\it objectively} in exactly the same 
tenseless or atemporal sense in which what exists at the time $t$ exists.)

Since nothing in the physical world corresponds to the distinction between what exists 
{\it now} and what does {\it not yet} exist, the future (relative to a time $t$) is as closed 
as the past. If a measurement of the value possessed by $Q$ at the time $t$ is 
successfully performed at the time $t'\geq t$, and the result is~$q$, then it always has 
been and always will be true that the value of $Q$ at the time $t$ is~$q$. There is 
nothing physically open about this. From the point of view of physics, the outcomes of 
performed measurements are always ``fixed and settled''~\cite{Stapp}. Nothing 
objectively changes at the time~$t$ (unless $t=t'$). What objectively changes, at the time 
$t'$, is that subsequently there obtains an actual state of affairs from which the value of 
$Q$ at the time $t$ can be inferred. The causal interconnectedness of the classical 
domain ensures that the possibility (in principle) of inferring the value of $Q$ at the 
time $t$ will persist. (See Sec.~\ref{cd} below. In measurement theory this is usually 
referred to as the creation of a record~\cite{Peres00}.) The misconception that the value 
of $Q$ at the time $t$ did not exist until it was observed or became inferable has to be 
seen for what it is: a naive objectification or reification of our ignorance, which ceases to 
exist, in principle at the time~$t'$, and in actual fact when we look at the pointer. What 
{\it is} (and always has been, and always will be) objectively open is the results of {\it 
unperformed} measurements.

\section{\large QUANTUM MECHANICS AND\\
CAUSALITY}

Physics concerns spatiotemporal regularities. Classical physics concerns deterministic 
regularities that permit us to make (i)~definite predictions on the basis of initial 
positions in phase space, (ii) definite retrodictions on the basis of final positions in phase 
space, and (iii)~definite inferences of intermediate states on the basis of initial and final 
positions in configuration space. Quantum physics concerns statistical regularities that 
permit us to assign (i)~prior probabilities on the basis of earlier property-indicating 
facts according to the Born rule, (ii)~posterior probabilities on the basis of later 
property-indicating facts according to the same rule, and (iii)~time-symmetric 
probabilities on the basis of earlier and later property-indicating facts according to the 
ABL~rule~\cite{UMABL}, so named after Aharonov, Bergmann, and 
Lebowitz~\cite{ABL64}.

The deterministic regularities of classical physics lend themselves to a causal 
interpretation, according to which the observable regularities are due to unobservable 
causal strings by means of which earlier events necessitate later events. Since the 
time-symmetric laws of classical physics provide no objective foundation for it, this 
time-asymmetric interpretation can be nothing but an anthropomorphic projection, into 
the time-symmetric world of classical physics, of the successive perspective of a conscious 
agent. Causality lies in the eye of the beholder~\cite{Hume}. It is our way of 
interpreting events, not a feature of the events in themselves. At best it is a secondary 
quality like pink or turquoise~\cite{MenziesPrice}.

The statistical regularities of quantum physics do not admit of this anthropomorphic 
projection. To begin with, measurement outcomes are {\it causal primaries}. They are 
value-indicating states of affairs that are {\it not} necessitated by antecedent causes. To 
see this, recall from Sec.~\ref{qmc} that all quantum-mechanical probability assignments 
involve the assumption that the value of a specific observable at a specific time is 
indicated by an actual state of affairs. Quantum mechanics never predicts that a 
measurement will take place, nor when one will take place. And if QM is the 
fundamental and complete theoretical framework that most of us believe it is, these 
things cannot be predicted because there is nothing that {\it necessitates} the existence of 
a value-indicating state of affairs.

I do not mean to say that in general nothing causes a measurement to yield this 
particular outcome rather than that. Unless hidden variables are postulated, this is a 
triviality. What I am saying is that nothing ever {\it causes} a measurement to take 
place. A clear distinction between two kinds of probability must be maintained. In the 
context of a position measurement using an array of detectors, this is the distinction 
between the probability {\it that} a detector will respond (no matter which) and the 
probability that a specific detector will respond {\it given} that any one detector will 
respond. The latter probability is the one that quantum mechanics is concerned with. 
The former probability can be measured (for instance, by using similar detectors in 
series), but it cannot be calculated from first principles, for essentially the same reason 
that a fundamental coupling constant cannot be so calculated~\cite{Mohrhoff00}.

If anything is causally determined, it can only be the probabilities associated with 
measurement outcomes. But the dependence of quantum-mechanical probability 
assignments on facts also does not admit of a causal interpretation, for at least two 
reasons. For one, there isn't just one way of assigning probabilities. If probability 
assignments are to be of any use, they have to be based on facts. But the probability 
$p\,(Q{=}q,t)$ of the result $q$ of a measurement of $Q$ at the time $t$ can be assigned 
on the basis of several different (sets of) facts. After all, probabilities are nothing but 
best guesses {\it given} the facts that are taken into account. This is true even if Nature 
herself tells us (via the laws of QM) what probabilities we should assign, for Nature does 
not tell us which facts should be taken into account. This choice is left to us. Stapp's 
belief that this choice too is dictated by Nature---that the probability $p\,(Q{=}q,t)$ is 
uniquely determined by a unique past (relative to $t$)---is an error.

Another reason why quantum-mechanical probabilities cannot be causally explained is 
that a best guess is not the kind of thing that admits of a causal explanation. While a 
best guess obviously depends on a {\it chosen} set of facts, it cannot be construed as an 
objective propensity that is causally determined by a {\it unique} set of facts, without 
committing some of the fallacies pointed out above.

The ontological dependence of all possessed properties on property-indicating states of 
affairs, combined with the fact that such states of affairs are causal primaries, implies 
that at a fundamental level {\it nothing in the physical world is necessitated by 
antecedent causes.} The acausal foundation of the physical world ought to be regarded 
as one of the most important discoveries of modern science, rather than deplored as the 
``Copenhagen renunciation'' of all ``attempts to understand physical 
reality''~\cite{Stapp}. As the following sections will show, the impossibility of causally 
construing the quantum-mechanical correlations has significant ontological implications. 
It does not entail that we must ``renounce for all time the aim of trying to understand 
the world in which we live''~\cite{Stapp}.

\section{\large OBJECTIVE PROBABILITIES}
\label{op}
N. David Mermin believes that all the mysteries of quantum mechanics can be reduced 
to the single puzzle posed by the existence of objective probabilities~\cite{Mermin98}. I 
concur. The objective probabilities Mermin has in mind, however, are not the objectified 
evolving probabilities Stapp calls ``propensities.'' Probabilities qualify as objective if 
and only if 
they are assigned on the basis of all relevant facts, so that there is nothing to be ignorant 
of. The probabilities of classical statistical physics are always subjective; they make up 
for facts that we ignore. The prior Born probabilities that we assign to the possible 
results of performed measurements are equally subjective, inasmuch as we assign them 
without taking the actual results into account.

The term ``objective probability'' has two natural definitions. When there {\it are} 
actual states of affairs having a bearing on the probability $p\,(Q{=}q,t)$, this 
probability is objective only if all of them are taken into account, regardless of whether 
they obtain before or after the time~$t$. In this case $p\,(Q{=}q,t)$ is given by the ABL 
rule~\cite{Mohrhoff00,UMABL}.

The second definition is appropriate when there aren't any relevant facts. This is the 
case when we are dealing with the stationary probability measures that solve the 
time-independent Schr\"odinger equation, rather than with probability measures that 
depend on initial and/or final conditions imposed on the time-dependent Schr\"odinger 
equation. The Born probabilities we obtain from stationary states are objective just in 
case they are counterfactually assigned.

What is the ontological significance of objective Born probabilities? The pertinent issue 
is the stability of matter. When Bohr wrote his doctoral thesis on the electron theory of 
metals, he became fascinated by its instabilities. They suggested to him a new type of 
stabilizing force, one fundamentally different from those familiar in classical physics. 
Transferring his postdoctoral attention from metals to Rutherford's atom, Bohr realized 
that this too ought to be unstable, and by imposing his well-known quantum conditions 
he took the first step toward an understanding of the force that stabilizes matter. When 
the mature theory arrived twelve years later, it transpired that this ``force'' hinges on 
the fuzziness of internal spatial relations. What ``fluffs out'' matter is the indefiniteness 
of the relative positions of its constituents. Though today this fact is known to 
every physicist, few seem to attach to it the importance that it deserves.

The proper way of dealing with fuzzy values is to make counterfactual probability 
assignments. If a quantity is said to have an ``indefinite value,'' what is really intended is 
that it does not actually have a value (inasmuch as the value is not measured) but that it 
{\it would} have a value if this {\it were} indicated, and that at least two possible values 
are associated with positive probabilities. (The counterfactuality cannot be eliminated, 
though it may be shifted from measurements to fuzzy values: If measurements of an 
observable $Q$ {\it are} successfully performed on an ensemble of identically prepared 
systems, and if the results have positive dispersion, the value of $Q$ {\it would be} fuzzy 
with regard to an individual system $S$ if the measurement {\it were not} performed on 
$S$.)

The objective probabilities associated with the results of unperformed measurements 
thus are the formal expression of an {\it objective indefiniteness}. The extrinsic nature of 
the values of quantum-mechanical observables follows directly from this indefiniteness. 
It is a straightforward consequence of an objective fuzziness in the world in which we 
live. Here is how: The proper expression of this fuzziness, as we just saw, involves 
counterfactual probability assignments. These assignments are based on the observable 
statistical correlations that QM encapsulates. We are therefore dealing with conditional 
probability assignments whose antecedents may or may not be true. We therefore need a 
criterion for when an antecedent is true, and this consists in the existence of a 
value-indicating fact.

The extrinsic nature of possessed values is not confined to atoms and suchlike. As Peres 
and Zurek~\cite{PerZur} rightly insist, ``[t]here is nothing in quantum theory making it 
applicable to three atoms and inapplicable to $10^{23}$.'' Although the moon isn't only 
there when somebody looks~\cite{Mermoon}, it is there only because of the myriad of 
facts that betoken 
its presence. If there weren't any actual state of affairs from which its position could be 
inferred, it wouldn't have a position, or else its position wouldn't have a value. (There is 
no need for anyone to actually carry out the inference.) This seems to entail a vicious 
regress, which at first blush looks like just another version of von Neumann's 
catastrophe of infinite regression.

The positions of detectors are extrinsic too. They are what they are only because of the 
facts that indicate what they are. This seems to require the existence of detector 
detectors indicating the positions of detectors, which seems to require the existence of 
detectors indicating the positions of detector detectors, and so on {\it ad infinitum}. 
Generally speaking, the contingent properties of things ``dangle'' ontologically from 
what is the case in the rest of the world. Yet what is the case there can only be described 
by describing material objects, and the properties of such objects too ``dangle'' from the 
goings-on in the rest of the world. This seems to send us chasing the ultimate 
property-indicating facts in never-ending circles.

So were does the buck stop? As it stands, the question is ill posed. There are no valueless 
positions in search of value-indicating facts. There are facts, and there are statistical 
correlations among value-indicating facts. These correlations warrant inferences to the 
existence of objects with properties that have indefinite values (in the sense explained 
above), and they also warrant the interpretation of the statistically correlated facts as 
indicating possessed values. The genuine core of the ``measurement problem'' is this: 
Value-indicating facts are actual states of affairs. Facts are by definition factual {\it per 
se}. Yet a state of affairs involves objects and their properties, and the properties of 
objects are extrinsic; their possession is not factual {\it per se}. So what justifies our 
treating the value-indicating properties of value-indicating things as if they were 
intrinsic (the opposite of extrinsic)? The answer will be given in the following section.

\section{\large THE CLASSICAL DOMAIN}
\label{cd}
The positional indefiniteness of a material object $O$ evinces itself through the 
unpredictability of the results of position measurements performed on~$O$. Evidence of 
the corresponding statistical dispersion requires the existence of detectors with sensitive 
regions that are small and localized enough to probe the range of values over which 
$O$'s position is distributed. (A detector is any object capable of indicating the presence 
of another object in a particular region.) If there are no such detectors, the 
indefiniteness of $O$'s position cannot evince itself. But detectors with sharper positions 
and sufficiently small sensitive regions cannot exist for all detectable objects. Since no 
relative position is absolutely sharp, there is a finite limit to the sharpness of the 
positions of material objects, and there is a finite limit to the spatial resolution of 
actually existing detectors. Hence there are objects whose positions are the sharpest in 
existence. These never evince their indefiniteness through unpredictable 
position-indicating facts. Such objects deserve to be designated ``macroscopic.'' We 
cannot be certain that a given object qualifies as macroscopic, inasmuch as not all 
matters of fact about its whereabouts are accessible to us, but we can be certain that 
macroscopic objects exist.

If the positional indefiniteness of a macroscopic object never evinces itself through 
unpredictable position-indicating events---the occasional unpredictability of the position 
of a macroscopic pointer reveals the indefiniteness of a property of a different object, not 
the indefiniteness of the position of the pointer---then it is legitimate to ignore the 
positional indefiniteness of macroscopic objects. We can associate such objects with 
classical trajectories, provided we understand the claims involved. It is not claimed that 
macroscopic objects have exact positions. It is only claimed that since their positions are 
the most definite in existence, the indefiniteness of their positions never shows up in the 
realm of facts. If we make the assumption that macroscopic objects follow definite 
trajectories, we will never see this assumption contradicted by facts. But if it is 
legitimate to ignore the positional indefiniteness of macroscopic objects, it is also 
legitimate to treat the positions of macroscopic objects as intrinsic.

The step from acknowledging the extrinsic nature of all contingent properties to treating 
the positions of macroscopic objects as intrinsic is of the same nature as the step from 
acknowledging the purely correlative character of classical laws of motion to the use of 
causal language. Macroscopic objects evolve predictably in the sense that every time the 
position of such an object is indicated, its value is consistent with all predictions made 
on the basis of (i)~all past indicated properties and (ii)~the classical laws of motion. (As 
mentioned above, there is one exception: Whenever the position of such an object serves 
to indicate an unpredictable property in the quantum domain, it is itself not 
predictable.) This makes it possible to think of the positions of macroscopic objects as 
forming a self-contained system of positions that ``dangle'' causally from each other, and 
this makes it possible to disregard that in reality they ``dangle'' ontologically from 
position-indicating facts. Predictability warrants the applicability of causal concepts, 
and the applicability of causal concepts to macroscopic objects warrants treating their 
positions as intrinsic.

Causality and intrinsic properties therefore stand and fall together. Where the extrinsic 
nature of properties cannot be ignored, causal concepts cannot be applied. While 
correlations that are not manifestly probabilistic (like those between the successive 
positions of macroscopic objects) can be embellished with causal stories, in the quantum 
domain causal concepts are out of place. We can impose them on the classical domain 
with some measure of consistency, although this entails the use of a wrong criterion: 
Temporal precedence takes the place of causal independence as the criterion that 
distinguishes causes from effects. But when we deal with correlations that are manifestly 
probabilistic, projecting our agent causality into the physical world does not work. 
Trying to causally explain these correlations is putting the cart in front of the horse. It is 
the statistical correlations that explain why causal explanations work to the extent they 
do. They work in the classical domain where statistical variations are not in evidence. If 
we go beyond this domain, we realize that all correlations are essentially statistical, even 
where statistical variations are not in evidence, and that causality is a function of 
psychology rather than a physical concept.

\section{\large THE SPATIOTEMPORAL\\
DIFFERENTIATION OF REALITY}
\label{stdr}
According to Richard Feynman, the mother of all quantum effects is the strange 
behavior of electrons in two-slit experiments~\cite{Feynmanetal65}. If nothing indicates 
the slit taken by an electron then this electron goes through both slits without going 
through a particular slit and without having parts that go through different slits. The 
bafflement caused by this behavior is symptomatic of a mismatch between the spatial 
aspect of the physical world and the way in which we all tend to think about space, for 
psychological and neurophysiological reasons~\cite{UMCCP,UMBCCP}.

We tend to think of space as a set of points cardinally equal to the set \RRR\ of triplets 
of real numbers, or else we tend to think of space as an extended thing that it is 
inherently divided into mutually disjoint regions (philosophically speaking, an extended 
substance with intrinsic parts), and we consider it legitimate to mathematically 
represent this thing by the set \RRR. (If it's intrinsically divided, then it's divided into 
infinitesimal regions, and then it seems OK to represent these regions by the elements of 
\RRR.)

QM is trying to tell us otherwise. The only positions in existence are (i)~the (not 
manifestly fuzzy) positions of macroscopic objects and (ii)~the positions possessed by 
objects in the quantum domain. The latter are defined by the sensitive regions of 
macroscopic detectors, are always finite in extent, and are possessed only when 
indicated. The proper way of thinking, speaking, or writing about a ``region of space,'' 
therefore, is to never let the logical or grammatical subject of a sentence refer to it. 
Regions of space are not things that exist by themselves, nor are they parts of a thing 
that exists by itself. Since they exist only as properties of material things, only predicates 
of sentences about material things should refer to them.

If they were things, the regions defined by the two slits---let's call them $L$ and 
$R$---would be distinct, self-existent parts of space. An object that is in the union 
$U=L\cup R$ of two 
distinct, self-existent parts of space, either is in $L$, or is in $R$, or is divided by the 
distinctness of $L$ and $R$ into two parts, one in $L$ and one in $R$. Since electrons in 
two-slit experiments can go through $U$ without going through either $L$ or $R$ and 
without being divided into parts that go through different slits, the two slits cannot be 
things. $L$ and $R$ are properties that are possessed if and only if their possession is 
indicated. If they are not possessed (because they are not indicated) then they do not 
exist. But if they do not exist, they obviously cannot compel electrons to ``choose'' 
between them.

A position measurement performed on $O$ at a time $t$ with $N$ detectors $D_i$ 
(sensitive regions $R_i$) answers $N$ yes-no questions. It yields truth values (``true'' or 
``false'') for $N$ propositions of the form ${\bf p}_i=$``$O$~is inside $R_i$ at the 
time~$t$.'' Where $O$ is concerned, the world at the time $t$ is spatially differentiated 
into $N$ finite regions. They exist for $O$ because the propositions ${\bf p}_i$ are either 
true or false, and these propositions are either true or false because their truth values are 
indicated. (If their truth values are not indicated, they are neither true nor false but 
meaningless.)

That we can treat the positions of macroscopic objects as intrinsic, for reasons and 
subject to qualifications stated in the previous section, does not change the fact that at 
bottom they too are extrinsic. While the whereabouts of macroscopic objects are 
abundantly and redundantly indicated, they are never indicated with absolute precision. 
Hence even for a macroscopic object $O$ the world at any given time $t$ is only finitely 
differentiated spacewise (that is, no finite region $R$ is differentiated into infinitely 
many regions $R_i$ such that truth values exist for all propositions ${\bf p}_i$).

The {\it finite} spatial differentiation of reality is one of the most significant ontological 
implications of QM~\cite{Mohrhoff00,UMCCP}. It is as inconsistent with the 
field-theoretic notion that physical properties are instantiated by the ``points of 
space''%
\footnote{``A field theory in physics is a theory which associates certain properties with 
every point of space and time.''---M. Redhead~\cite{RedheadQFT}.}
as special relativity is with the notion of absolute simultaneity. The world is created 
top-down, by a finite process of differentiation that stops short of an infinite spatial 
differentiation, rather than built bottom-up, on an infinitely and intrinsically 
differentiated space, out of locally instantiated physical properties. There are no points 
on which to build such a world. An infinitely and intrinsically differentiated space, such 
as \RRR\ is commonly supposed to represent, exists nowhere but in our thoughts. We 
may think of the trajectories of macroscopic objects as the paths of average (expected) 
positions, but the fuzziness implied by this way of thinking exists solely in our 
imagination. It corresponds to nothing in the physical world because it exists only in 
relation to an unrealized degree of spatial differentiation---it exists only in relation to an 
imagined backdrop that is more differentiated spacewise than is the physical world.

What is true of the world's spatial aspect is equally true of its temporal aspect. There is 
no such thing as an intrinsically and infinitely differentiated time. What is temporally 
differentiated is physical systems, and every physical system is temporally differentiated 
only to the extent that it has distinct successive states, in the common-language sense of 
``state'' that connotes possessed properties. The world's limited temporal differentiation 
is a direct consequence of its limited spatial differentiation. Because the world is only 
finitely differentiated spacewise, no physical system can have an infinite number of 
distinct states in a finite time span~$T$. Therefore a macroscopic clock (usually 
indicating time by some macroscopic pointer) can indicate no more than a finite number 
of distinct times during~$T$, and this means that there exist no more than a finite 
number of such times during~$T$.

Consider a system $S$ to which, on the basis of its factually warranted properties at the 
indicated clock times $t_1$ and $t_2$, the respective Born probability measures 
$\ket{\psi_1(t)}$ and $\ket{\psi_2(t)}$ can be assigned. If there isn't any fact that 
indicates what $S$ is like in the meantime then there isn't anything that $S$ is like in 
the meantime. Where $S$ is concerned, there isn't any state (in the common-language 
sense of the word) that obtains in the meantime, let alone an evolving instantaneous 
state. ``[T]here is no interpolating wave function giving the `state of the system' between 
measurements''~\cite{Peres84}. Not only is there no state that obtains in the meantime 
but also there is no meantime. And so there isn't any time at which propensities can be 
attributed to~$S$. Times, like properties, supervene on the facts. Not only the positions 
of things but also the times at which they are possessed are extrinsic. The times that exist 
for $S$ are the factually warranted times at which $S$ possesses factually warranted 
properties.

If there isn't any matter of fact about what $S$ is like in the meantime, we can say that 
$S$ has {\it changed} from an object having properties that warrant assigning 
$\ket{\psi_1(t)}$ to an object having properties that warrant assigning 
$\ket{\psi_2(t)}$---but {\it only} in the sense that at the time $t_1$ the system has the 
former properties and at $t_2$ it has the latter properties. The change of $S$ {\it 
consists} in the difference between the properties it possesses at $t_1$ and the properties 
it possesses at~$t_2$. Where $S$ is concerned, this is all the change that occurs. Nothing 
can be said about the meantime, not just because in the meantime $S$ lacks 
properties, but because there isn't any meantime. (Much the same is true of positions 
between material objects. If a position somewhere between two material objects is not 
possessed by another material object, it does not exist.)

Stapp's 1972 interpretation of the Copenhagen interpretation combines one correct idea 
with two erroneous notions~\cite{StappCI}:
\bq
The rejection of classical theory in favor of quantum theory represents, in essence, the 
rejection of the idea that external reality resides in, or inheres in, a space-time 
continuum. It signalizes the recognition that `space,' like color, lies in the mind of the 
beholder.

The principal difficulty in understanding quantum theory lies in the fact that its 
completeness is incompatible with [the] external existence of the space-time continuum 
of classical physics.

The theoretical structure did not extend down and anchor itself on fundamental 
microscopic space-time realities. Instead it turned back and anchored itself in the 
concrete sense realities that form the basis of social life.
\eq
While it is correct that QM is incompatible with the space-time continuum of classical 
physics, the conclusion that space ``lies in the mind of the beholder'' is a {\it non 
sequitur}, and so is the notion that the theoretical structure of QM is anchored in 
``concrete sense realities.'' What is inconsistent with QM is the existence of an 
intrinsically and infinitely differentiated space-time continuum. Neither space nor time 
is a world constituent that exists independently of matter. Therefore neither can be {\it 
intrinsically} differentiated. Space and time are modes of differentiation. The objective 
world {\it is} in possession of spatial and temporal aspects, but it is only {\it finitely} 
differentiated spacewise and timewise. Again, QM presupposes measurements, but 
measurements {\it qua} value-indicating facts, not measurements {\it qua} ``concrete 
sense realities.''

In his present theory, Stapp rejects the one correct ingredient in his 1972 interpretation 
and postulates, as the arena for a local dynamical process satisfying the principle of local 
causality, the infinitely and intrinsically differentiated space-time continuum of 
relativistic quantum field theory:
\bq
The evolution of the physical universe involves three related processes. The first is the 
deterministic evolution of the state of the physical universe. It is controlled by the 
Schr\"odinger equation of relativistic quantum field theory. This process is a local 
dynamical process, with all the causal connections arising solely from interactions 
between neighboring localized microscopic elements.
\eq
As pointed out earlier in this section, these field theoretic notions are as inconsistent 
with the finite differentiation of the objective world implied by QM as the notion of 
absolute simultaneity is with special relativity.

\section{\large QUANTUM MECHANICS AND\\
MIND--BRAIN INTERACTION}

On the basis of von Neumann's discordant postulates---a classical space-time continuum, 
a dynamical process subject to local causality, instantaneous collapse due to the injection 
of ``information associated with a subjective perception by some observing system into 
the objective state of the universe''~\cite{Stapp}---Stapp formulates a theory of 
mind--brain interaction in which choices are claimed to play a crucial role: ``The basic 
building blocks 
of the new conception of nature are\ldots choices of questions and answers.'' We freely 
choose the questions, and Nature freely chooses the answers, within the constraints 
imposed by the statistical laws of QM.

Like Bohr, Stapp attributes to experimenters the freedom to choose between 
complementary experimental arrangements: ``the choice of which question will be put to 
nature\ldots is not governed by the physical laws of contemporary physics.'' Here I 
agree. The eventual physical effect of such a choice---the experimental setup that is 
actually in place---is not determined by any of the presently known physical laws, nor is 
the initial physical effect, which causes, in accordance with the neuroscience of motor control, the actions that lead to the eventual effect.

Just as a specific event in classical physics always leads, under identical conditions, to 
the same effect, so a specific, causally efficacious plan of action always leads to a specific, 
causally determined course of action. It is precisely because there is no indeterminacy in 
the correlations between causes and effects that we can speak of 
``causes'' and ``effects.'' Since causal concepts are applicable only to dispersion-free 
correlations, quantum indeterminacy can play no mediating role in mental causation. 
The freedom to choose is a classical phenomenon. The difference between physical 
causation and mental causation is that the causes of the former, like their effects, belong 
to the classical domain, while the causes of the latter are not to be found in either 
physical domain.

The description of the physical effects of mental causes cannot differ from the 
description of the physical effects of physical causes. The effects of mental causation 
must be capable of being represented by the same mathematical constructs as the effects 
of physical causation---that is, by one of the classical force fields. As has been shown 
elsewhere~\cite{UMIEC,UMPI}, the relevant field is the electromagnetic four-vector 
potential. Where this is only physically determined, it is determined (up to gauge 
transformations) in conformity with Maxwell's equations. Where it includes the effects 
of causally efficacious mental events, it is no longer so determined.

Stapp asserts that ``[a]ccording to the principles of classical physical theory, 
consciousness makes no difference in behavior: all behavior is determined by 
microscopic causation.'' In point of fact, this is so according to certain metaphysical 
doctrines, not according to the principles of any physical theory. Consciousness can make a difference, although not without infringing physical laws~\cite{UMIEC,UMPI}. Like J.C. 
Eccles~\cite{Eccles}, Stapp appears to hope that QM will allow the mind to be causally 
efficacious without infringing physical laws. While Eccles tried, 
unsuccessfully~\cite{UMIEC,UMPI}, to exploit quantum-mechanical indeterminism as a 
loophole through which mind can act on matter without ``violating'' the laws of physics, 
Stapp tries to explain the freedom of the experimenter by the freedom of the 
experimenter's mind to pay or not to pay attention. His argument involves the following 
seven steps.
\medno
(1) Environment-induced decoherence (EID)~\cite{Zurek93} ``creates a powerful 
tendency for the brain to transform almost instantly into an ensemble of components, 
each of which is very similar to an {\it entire classically-described brain}''~\cite{Stapp}.
\medno
Two comments. First, the transformation to which Stapp refers is not a transformation 
of the brain but the ``transformation'' of a probability measure associated with one time 
into a probability measure associated with another, slightly later time. If the prior 
probability measure associated with the brain and the time $t$ is a coherent 
superposition
\bdm
\sum_i\sum_k c^*_i c_k \ket i\bra k
\edm
or a mixture of such superpositions
\bdm
\sum_j \lambda_j\sum_i\sum_k c^*_{ji} c_{jk} \ket i\bra k,
\edm
the effect of EID is that the prior probability measure associated with the brain and 
a slightly later time $t'$ is approximately given by the mixture
\bdm
\sum_k c^*_k c_k \ket k\bra k
\edm
or the mixture (of mixtures)
\bdm
\sum_j \lambda_j\sum_k c^*_{jk} c_{jk} \ket k\bra k.
\edm
Second, if there is a way of making sense of the phrase ``classically described brain'' (CDB), the phrase refers to a brain the positional indefiniteness of whose material 
constituents is not evidenced by such position-indicating facts as are inconsistent with 
classical laws of motion (Sec.~\ref{cd}).
\medno
(2) Each instance of EID in the brain is preceded and brought into play by exocytosis, 
the release of the contents of a vesicle of neurotransmitter into the synaptic cleft.
\medno
The prior probability measure associated with the brain after exocytosis assigns 
significant probabilities to significantly different outcomes of position measurements 
that might be performed on some of the brain's material constituents. ``Significantly 
different'' is short for ``sufficiently different for the probability measure associated with 
the brain to be subject to EID.'' The net result, according to Stapp, is ``a quantum 
splitting of the brain into different classically describable components'' or elements, 
all of which must be regarded as real ``because interference between the different 
elements [is] in principle possible.''

In point of fact, the net result of exocytosis and EID is a mixed probability measure. As a 
macroscopic object, a CDB is associated with a probability measure that does not assign 
significant probabilities to significantly different outcomes of possible position 
measurements on its material constituents. For this reason we can switch from 
conditional assignments of probabilities to unconditional attributions of properties (that 
is, to attributions of {\it intrinsic} properties); we can talk facts. That is just why we can speak of a ``classically described'' brain. A ``mixture of CDBs,'' on the other hand, has no sensible translation into the classical language of objects and facts. It is neither an object nor an actual state of affairs but a probability measure pure and simple. 

If there isn't any matter of fact about which component of a mixture 
exists, no component exists. None of them can be ``regarded as real.'' This is 
the reason why re-interference remains a theoretical possibility. If nothing in the 
decoherence-inducing environment indicates a particular component, coherence can in 
principle be restored. Since correlations between the respective probability measures of 
the environment and the brain are necessary but not sufficient for the existence of a 
component-indicating fact, the probability measure associated with the brain at a later 
time could in principle be the initial pure measure, assuming that the initial measure was 
pure. Whether this is possible under conditions in which a living brain can exist, is a 
different matter. It stands to reason that under such conditions the 
decoherence-inducing environment intersects with the classical domain. If so, the 
existing component will be 
indicated, in which case the restoration of coherence is ruled out.

A decoherent mixture can be objective in the sense that it specifies objective 
probabilities. (I agree with Stapp that decoherence is not sufficient for the 
transformation of objective probabilities into subjective ones.) As explained in 
Sec.~\ref{op}, objective probabilities are associated with the possible results of 
unperformed measurements. An objective probability measure is the proper, 
counterfactual expression of an objective indefiniteness. If nothing indicates the slit 
taken by an electron, an objective probability of~$1/2$ can be assigned to the possibility 
that the electron has gone through the left (right) slit. In this case saying that the 
electron went through the left (right) slit is neither true nor false. It is meaningless, for 
the distinction we make between these alternatives is a distinction that Nature does not 
make; it corresponds to nothing in the physical world; it exists solely in our minds.

In view of the unavoidable intersection between the decoherence-inducing environment 
and the classical domain, a mixture of {\it living} CDBs can only be a subjective 
probability measure, arising from an incomplete knowledge of the relevant facts. It 
cannot represent the objective brain. Let us assume, nevertheless, that a ``mixture of 
CDBs'' is an objective probability measure. (Only in this case does it make sense to keep 
looking for a process that changes ``and'' into ``or,'' or objective probabilities into 
subjective ones, and in which consciousness can play a causal role.) Then this objective 
probability measure is the formal expression of an objective indefiniteness, and the 
distinctions that we make between its components are distinctions that Nature does not 
make; they correspond to nothing in the physical world. Hence even if a ``mixture of 
CDBs'' were an objective probability measure, its components could not be regarded as 
being both real and distinct from each other.
\medno
(3) ``[D]uring an interval of conscious thinking, the brain changes by an alternation 
between two processes.'' Having generated ``by a local deterministic mechanical rule'' a 
``profusion'' of ``separate, but equally real, quasi-classical branches,'' the ``individual 
physical system associated with a mental event'' undergoes a change by which it ``is 
brought into alignment with the content of that mental event.'' The physical aspect of 
this second process ``chops off all branches that are incompatible with the associated 
psychical aspect.''

``[I]f the psychical event is the experiencing of some feature of the physical world, then 
the associated physical event'' updates ``the brain's representation of that aspect of the 
physical world. This updating of the (quantum) brain is achieved by discarding from the 
ensemble of quasi-classical brain states all those branches in which the brain's 
representation of the physical world is incompatible with the information content of the 
psychical event''~\cite{Stapp}.
\medno
If the psychical event is the intention to execute a particular plan of action, then the 
associated physical event discards the branches associated with plans of action that are 
incompatible with the experienced intention. (Stapp assumes that ``the purely 
mechanical evolution of the state of the brain in accordance with the Schr\"odinger 
equation will normally cause the brain to evolve into a growing ensemble of alternative 
branches, each of which is essentially an entire classically described brain that specifies a 
possible plan of action.'')

Holding the definiteness of perceptions responsible for the quantum-mechanical 
``reduction process'' permits Stapp to pass from perceptions to other mental contents, 
and to similarly empower volitions: The definiteness of intentions, like that of 
perceptions, entails reductions. As yet, however, there is no room for free choices. 
Consciousness (a psychical aspect of perceptual and/or volitional character) is associated 
with all branches, and reduction occurs {\it automatically} whenever the mixture of 
branches becomes inconsistent with the definiteness of mental contents. To make room 
for free choices, Stapp introduces the metaphor of the experimenter as interrogator of 
Nature. Making experiments is asking {\it questions}, and getting results is receiving {\it 
answers} from Nature. The experimenter has the freedom to choose which experiments 
to perform when, and Nature has the freedom to chose the results.
\medno
(4) ``The central roles in quantum theory of these discrete choices---the choices of which 
questions will be put to nature, and which answer nature delivers---makes quantum 
theory a theory of discrete events\ldots. Each of these quantum events involves'' (i)~``a 
choice of a Yes-No question by the mind--brain system'' and (ii)~``a choice by Nature of 
an answer, either Yes or No, to that question. \ldots the freedom to choose which 
questions are put to nature, and when they are asked, allows mind to influence the 
behaviour of the brain''~\cite{Stapp}.
\medno
Within the Copenhagen framework, the interrogation of Nature by human 
experimenters is a fitting metaphor for a well-defined scenario. To choose a question is 
to decide on a specific experimental arrangement, and to choose a time is to decide when 
to perform the experiment. The mind, however, doesn't experiment with the brain. To 
save his metaphor, Stapp therefore needs to imbue it with a new sense. In what sense 
does the mind put 
questions to the brain, without the intervention of an apparatus? How does it choose its 
questions and the times to ask them? In Stapp's opinion, {\it attention} holds the key.
\medno
(5) ``Asking a question about something is closely connected to focussing one's attention 
on it. Attending to something is the act of directing  one's mental power to some task. 
This task might be to update one's representation of some feature of the surrounding 
world, or to plan or execute some other sort of mental or physical action.'' 
\medno
Neurobiological data suggest that the world's neural representation contains far more 
information than its conscious mental representation, and this suggests that attention 
plays a crucial role in the relation between the two representations. Sudden changes in 
the visual field can not only {\it draw} attention, and thereby update the mental 
representation, but also draw it {\it away}, and thereby obliterate features of this 
representation~\cite{Cheshire}. (The updating of the mental representation that would 
normally follow a change in the neural representation, may be prevented by focusing 
attention on a different part of the visual field.) When it is not drawn willy-nilly, 
attention seems to be capable of being directed freely. According to William James, 
whom Stapp quotes approvingly, the power to direct our attention is limited to choices 
between keeping it focused on whatever has captured it or allowing it to be captured by 
something else:
\bq
[T]he whole drama of the voluntary life hinges on the attention, slightly more or slightly 
less, which rival motor ideas may receive.

The essential achievement of the will, in short, when it is most ``voluntary,'' is to attend 
to a difficult object and hold it fast before the mind. \ldots Effort of attention is thus the 
essential phenomenon of will~\cite{James}.
\eq
While this appears to be both plausible and consistent with the neurobiological data, it 
has nothing to do with probabilistic reductions of mixtures. The questions the mind can 
put to the brain, by choosing where to fix its attention, are always compatible, for the 
mind does not need to choose between mutually incompatible experimental 
arrangements. The relations between mental contents and neural states, in which 
attention appears to play a significant role, are therefore relations between the mind and 
a CDB, which belongs to the definite domain of intrinsic properties. It has nothing to do 
with the relations between this domain and the indefinite domain of extrinsic properties, 
with which QM is concerned. The intersection between QM and volition is empty.

I do not deny that a complete understanding of the brain must take into 
account the positional indefiniteness (and hence the quantum-mechanical nature) of the 
brain's constituents. But of this there can be no evidence in the correspondences 
between facts and their neural and mental representations. If 
attention is drawn to the highest bidder, the highest bidder is not a component of a 
mixture of CDBs but one among several neural events or activities competing for 
attention in one and the same CDB. And if attention roams freely (that is, if the mind 
can freely choose the questions it puts to the brain) then also there is nothing stochastic 
in the answers it receives. Each answer is determined by a definite aspect of a single 
CDB, rather than by the probabilistic reduction of a mixture of CDBs.
\medno
(6)~The interactions between the physical universe and the minds of observers have two 
aspects. The first ``is the role of the experimenter in choosing what to attend to; which 
aspect of nature he wants to probe; which question he wants to ask about the physical 
world\ldots. The second aspect is the recognition, or coming to know, the answer that 
nature returns''~\cite{Stapp}.
\medno
Here Stapp glosses over the disparity between physical experimentation and 
psychological attention by applying the same metaphor to both. In point of fact, 
experimenters do not choose what to attend to; they decide which experiment they will 
perform. It is one thing to choose between incompatible experimental arrangements with 
a view to obtaining information about a part of the physical world that cannot be 
obtained by simply looking at it. It is something else altogether to choose which directly 
accessible feature of a CDB to attend to. (``Directly accessible'' 
means ``without the intervention of any apparatus'' and thus ``without having to chose 
between incompatible setups.'') The necessity of a choice exists for totally different 
reasons, namely, in one case, the impossibility of simultaneous answers to logically 
inconsistent questions%
\footnote{It is not only practically impossible but {\it logically} inconsistent to ask both 
(i)~whether an atom went through the left or through the right cavity {\it and} 
(ii)~whether the same atom went through both cavities in phase or out of phase. See my 
discussion~\cite{Mohrhoff00,UMESW} of the experiment of Englert, Scully, and 
Walther~\cite{ESW91,ESW94}.}
and, in the other case, the brain's limited processing capacity.

Stapp's transference, via the interrogator metaphor, of the non-Boolean structure of the 
lattice of possible experimental answers to the lattice of possible answers returned by the 
brain seems to entail that different plans of actions necessarily correspond to different 
components of a mixture of CDBs. If this were the case, we could never consciously 
weigh the pros and cons of different possible courses of action, for then different plans of 
action could not coexist in the same mind, given that consciousness of a plan of action is 
what reduces the mixture to one of its components. To be consistent with the 
introspectively evident coexistence of alternative plans of action in the same mind, Stapp 
would have to allow either that different courses of actions can coexist in the same CDB 
or that we can be conscious of different components of a mixture of CDBs. But in the 
former case the choice between alternative plans of action cannot be linked to the 
reduction of a mixture, and in the latter Stapp's consciousness-based account of state 
reduction fails.

If the brain's limited processing capacity is the reason why attention is choosy then 
attention has to be regarded as being in part a neural process. This works in Stapp's 
favor inasmuch as he wants to account for the causal efficacy of the mind in 
quantum-mechanical terms. Since no single choice of an experiment can influence the 
result of another experiment, no single question posed by the mind can be causally 
efficacious. On this theoretical foundation it makes sense to let the physical system 
decide which questions will be asked when, and to restrict the mind's freedom to an 
influence on the rate at which questions are re-posed. This kind of influence {\it can} be 
causally efficacious, as the quantum Zeno effect~\cite{MiSu,ChiSu} demonstrates, and it 
agrees with the fact that attention is largely a neural (and hence neurally determined) 
process. But it also gives the quietus to the interrogator metaphor, for now Nature not 
only provides the answers but also asks the questions.

By gradually shifting both the content and the application of his metaphor, Stapp is able 
to make plausible a series of specious transitions. Once the metaphor has served its 
purpose, it is discarded. In the first of the quotations that follow, Stapp attributes to the 
mind the freedom to choose its questions. In the second, the mind chooses between 
whether or not to pose a question chosen by Nature, and it controls the rate of 
questioning. In the third, the mind only controls the rate at which questions chosen by 
Nature are repeated. And in the fourth, the mind's freedom is reduced to consenting to 
the rapid re-posing of questions chosen by Nature.
\bq
The only freedom in the theory---insofar as we leave Nature's choices alone---is the 
choice made by the individual about {\it which} question it will ask next, and {\it when} 
it will ask it. These are the only inputs of mind to the dynamics of the brain.

[T]he brain does most of the work, in a local mechanical way, and the mind, simply by 
means of choices between `Yes' or `No' options, and control over the {\it rate} at which 
questions are put to nature, merely gives top-level guidance.

Mental control comes in only through the option to rapidly pose [the] same question 
repeatedly, thus activating the Quantum Zeno Effect, which will tend to keep the state of 
the brain focussed on [a specific] plan of action\ldots.

[M]ind, by means of the limited effect of consenting to the rapid re-posing of the 
question already constructed and briefly presented by brain, can influence brain activity 
by causing this activity to stay focussed on the presented course of action.
\eq
According to Stapp, if we are very intent on a specific course of action, we must be very 
skeptical about its being the right course of action; we must keep asking ourselves 
rapidly, ``Shall I execute plan~X? Shall I execute plan~X? Shall I\ldots.'' This will make 
our executing plan~X highly probable. On the other hand, if we wish to abstain from a 
certain course of action, we must not keep asking ourselves whether it should be 
executed.
\medno
(7)~Stapp presents a simple dynamical model of mind--brain interaction in which ``the 
`best possible' question that could be asked by the individual at time $t$,'' given the 
state $S(t)$ of the universe at this time, is the question $P_{max}$ that maximizes ${\rm 
Tr}[P S(t)]$. This question is posed when the probability of a positive answer reaches a 
relative maximum.
\medno
Here Stapp introduces a new physical law, specifying which question Nature will ask 
herself next and when she will do so. Stapp thus effectively proposes a new theory, as 
different from standard QM as nonlinear adulterations of QM~\cite{GRW,Pearle}. The 
theory which he ends up formulating is completely different from the theory he 
initially professes to formulate, for in the beginning consciousness is responsible for state 
vector reductions, while in the end a new physical law is responsible---a law that in no 
wise depends on the presence of consciousness.

Thus in the end Stapp, like Eccles~\cite{Eccles}, fails to account for mental causation 
without implying ``violations'' of the laws of contemporary physics. Eccles did not 
introduce a new physical 
law, but he allowed the mind to load the quantum dice in the process of exocytosis, and 
this is tantamount to postulating mentally generated local modifications of physical 
laws~\cite{UMIEC,UMPI}. Stapp introduces a new physical law specifying which 
questions Nature asks herself, and when, and he allows the mind to modify the rates at 
which Nature interrogates herself. This, too, is tantamount to postulating mentally 
generated local modifications of a physical law---the very law Stapp himself has 
introduced.

\section{\large EPILOGUE}

Stapp asserts that his ``conceptualization of natural process arises\ldots directly from an 
examination of the mathematical structure injected into science by our study of the 
structure of the relationships between our experiences.'' It is a truism that science begins 
with relationships between experiences. But it does not end there. Science is driven by 
the desire to know how things {\it really} are. It owes its immense success in large 
measure to its powerful ``sustaining myth''~\cite{MerminSM}---the belief that this can 
be discovered. Knowing how things ``really'' are does not mean knowing how they are 
{\it in themselves}, independently of how they appear to us or how we conceive of them. 
By definition, that kind of knowledge is beyond our ken. What science aims to achieve is 
a strongly objective conception of reality (that is, a consistent way of thinking of 
experienced regularities as aspects of a world that does not depend on its being 
experienced). The very aim of science thus rules out the intersubjective or weakly 
objective conception of reality advocated by d'Espagnat~\cite{dERP}. It also rules out 
interpretations of standard QM that take the quantum state for more than a probability 
measure, inasmuch as such interpretations are inconsistent with strong 
objectivity~\cite{dERP,dEVR,dESEP,dEIHR}. It further rules out epistemic 
interpretations, including Stapp's.

If the aim of strong objectivity appears unattainable, it ought to be taken as a sign that 
we are making the wrong assumptions, and it ought to spur us on to ferret them out. 
The crucial assumption that stands in the way of a strongly objective conception of 
reality based on von Neumann's formulation of QM is the idea that the ``evolution'' of 
the physical state of the universe ``between events'' is ``a local dynamical process, with 
all the causal connections arising solely from interactions between neighboring localized 
microscopic elements''~\cite{Stapp}. For Stapp, reality is differentiated, both 
spacewise and timewise, into infinitesimal ``neighboring localized microscopic 
elements.'' If the objective physical state is ``an informational and dispositional 
substrate that carries both the information incorporated into it by the psychical realities, 
and certain dispositions for the occurrence of future psychical realities''~\cite{Stapp}, 
this locality assumption is completely gratuitous. The temporal resolution of the human 
visual system is in the millisecond range; its angular resolution is about an arc minute. 
Though higher resolutions can be achieved with the help of physical instruments, owing 
to intrinsic limits to the spatial and temporal magnifying power of such instruments, 
infinitesimal neighboring intervals or regions can never be distinguished. But if 
``psychical realities'' are only finitely differentiated spacewise and timewise, and if the 
physical state of the universe only ``carries information'' about past ``psychical 
realities'' and propensities for future ``psychical realities,'' then why should the physical 
universe be infinitely differentiated?

The recognition that the physical world is only finitely differentiated spacewise and 
timewise (Sec.~\ref{stdr}) clears the way for a rigorous objective distinction between the 
classical domain of intrinsic properties and the quantum domain of extrinsic properties 
(Sec.~\ref{cd}). It warrants the special status Bohr accorded to measurement 
outcomes---property-indicating facts---and makes it possible to establish an objective 
criterion for distinguishing measuring apparatuses from ``lesser things.'' Essential to 
attaining this objective was the realization that the attribution of factuality is beyond 
the scope of any theory. When the theory has done its part, we are left with the problem 
of assigning factuality. This problem has exactly one solution. The inexplicable factuality 
of facts belongs to those properties which, for all {\it quantitative} purposes, can be 
treated as intrinsic.

Instead of according a special status to measuring instruments, Stapp accords it to the 
neural correlates of mental states. Mental states evolve classically. The quantum brain 
does not, but the definiteness of mental states forces their neural correlates, and through 
them everything that is entangled with them, to behave in a 
classical manner. In order to turn his idea of a theory into a proper theory, Stapp would 
have to establish a criterion for distinguishing the neural correlates of mental states from 
less exalted aspects of the brain. If we are to reject the Copenhagen interpretation 
because it fails to establish a criterion for distinguishing measuring instruments, we 
should equally reject Stapp's theory, for it fails to establish a criterion for distinguishing 
the neural correlates of mental states. He does, however, venture the following 
conjecture:
\bq
This suggests to me that the physical correlates of the psychical realities will reside in the 
low frequency components of the coulomb part of the electromagnetic field. These are 
dominated by the so-called ``coherent states,'' which are known to be essentially classical 
in nature, and particularly robust\ldots. This would allow psychical realities\ldots to be 
present in the simplest life forms, and to predate life~\cite{Stapp}.
\eq
The attempt to identify the neural correlates of consciousness in physical terms appears 
to lead more or less inevitably to some form of panpsychism. This suggests to me that 
John Searle's comment on David Chalmers' functionalist account of 
consciousness~\cite{ChalmersCM} applies equally to Stapp's account: ``Of all the 
absurd results in Chalmers' book, panpsychism is the most absurd and provides us with 
a clue that something is radically wrong with the thesis that implies it''~\cite{Searle}.

\end{document}